\documentclass[12pt]{article}

\usepackage{amsmath,amssymb, amsthm}

\title{Classical signal viewpoint to bunching and anti-bunching}

\author{Andrei Khrennikov\\
International Center for Mathematical Modelling
\\in Physics and Cognitive Sciences\\
Linnaeus University,  V\"axj\"o, S-35195, Sweden}

\begin{document}

\maketitle

\begin{abstract} {\bf The similarity between classical wave mechanics and quantum mechanics (QM) played an important role 
in the development of QM (starting with works of De Broglie, Schr\"odinger, ``late Einstein'', Lamb, Lande,
Mandel, Marshall, Santos, Boyer, and many others). We present a new wave-type approach, so called prequantum classical statistical
field theory (PCSFT). PCSFT explores an analogy between some quantum phenomena and classical theory of random
fields. Quantum systems are interpreted as symbolic representations of such fields (not only for photons, cf. 
Lande and Lamb, but even for massive particles). All quantum averages and 
correlations (including composite systems in entangled states) can be represented as averages and correlations for classical 
random fields. In this paper PCSFT is used to provide a classical signal representation of bunching and anti-bunching. At least the  latter is typically considered as essentially  quantum (nonclassical) phenomenon. }
\end{abstract}

\section{Introduction}

In the quantum community, the viewpoint that quantum correlations cannot be reduced to correlations of classical random fields 
is very common, e.g. \cite{PL}, \cite{PL1}.   At the same time  various 
``prequantum models'', including models based on classical fields, have been developed, e.g. \cite{C2}--\cite{Kis}. These models 
reproduce some fundamental features of quantum mechanics (QM). Each model has its own advantages and disadvantages.
The classical field approach to QM is especially attractive. This approach reflects (see e.g.  \cite{PL1}) 
the ideas  of Planck, Mie, Schr\"odinger, De Broglie, and ``late 
Einstein''\footnote{In 1905 Einstein invented the quantum of light. However, since 1920th he started to work on a purely classical wave model
of physical reality, see e.g. \cite{EI}.}.  The {\it semiclassical approach} to quantum electrodynmaics
 and closely related to it, {\it stochastic 
electrodynamics} (SED),  see e.g. \cite{C2}-- \cite{Theo}, can be considered as the basic outputs the classical field approach to microphenomena.

A few years ago a new model of the classical field type, {\it prequantum classical statistical field theory} (PCSFT), was invented 
\cite{KH1}--\cite{KH14_a}. 
This is a purely field model, i.e., particles are totally excluded from consideration, cf. Mie, Einstein, Schr\"odinger. Not only photons, but 
all quantum systems, e.g. electrons, are described by classical random fields. Quantum measurements are interpreted as measurements of classical 
random signals. In some sense PCSFT is a model with hidden variables. At the moment we are not able to monitor prequantum signals, since
they fluctuate at the prequantum time scale which is essentially finer than the time scale of modern measurements \cite{KH1a}. However, one can expect 
that in future the prequantum time scale would be approached and it would be possible to forget about the quantum formalism 
(which provides only a very coarse description of microphenomena) and work  in the framework of the classical signal theory.\footnote{One can proceed in the inverse direction, i.e., create a quantum-like representation of the classical signal theory \cite{W};
 in particular, a coarse description of classical optical phenomena induces a (macroscopic) quantum-like model, see \cite{KH2}.}
We state again that signals are random.
Thus it is not clear whether Einstein's dream of re-establishing of determinism would come true. One of the main features of 
PCSFT is the assumption that randomness of prequantum fields corresponding to quantum systems is  
fundamentally coupled to a random background field
({\it ``fluctuations of vacuum''}).

We state again that the  {\it random background field plays a crucial role} in other classical field type models, especially in SED.
\footnote{Max Planck was the first who pointed out (in 1908 by debating with ``early Einstein'') 
that by taking into account a random background field one can escape quantum jumps.} As was emphasized, in the PCSFT-approach 
this is crucial that the background field is an irreducible component of random fields representing quantum systems. Opposite to e.g. 
SED, in PCSFT we could not distill the ``intrinsic field'' of e.g. an electron from the background field. Such 
an intrinsic field of a quantum system does not exist. It could not be distilled from fluctuations of vacuum. ``Filtering'' can be done only at the level of statistics of measurements, section \ref{LLKK}, see also  \cite{KH1b}-- \cite{KH1b2}.  This procedure is described by QM. Thus  QM probabilistic quantities do not coincide with classical ones. The former quantities are obtained from renormalization of the latter ones,  see (\ref{YY1a}), (\ref{YY2a}). Hence, quantum and classical models are connected in a rather nontrivial way.\footnote{Although we work
in the framework of quantum mechanics and not quantum field theory, renormalization connecting quantum mechanics and prequantum classical field theory, PCSFT, reminds a bit the QFT procedure for vacuum renormalization. For
some classical quantities (given by quadratic forms of classical fields corresponding to non trace class operators),
to get a quantum quantity, we have to subtract infinity from the corresponding classical quantity.}
And this connection is possible only under the presence 
of the background random field. This field produces additional correlations. In principle, one can consider the
background field as a source of ``nonlocal correlations''. However, this field is classical, so this is classical
wave ``nonlocality.''  The presence of the random background field (in combination with renormalization-like 
coupling between quantum and classical quantities) provides a possibility to create a classical probabilistic model
for anti-bunching. (It seems that such a possibility for bunching is not surprising). As the result of aforesaid 
features of PCSFT, we can escape problems related to existing proofs of impossibility of classical probabilistic 
description of anti-bunching, see, e.g., I. V. Volovich \cite{Volovich_B} for one of such proofs (presented in the rigorous mathematical framework) and based sub-Poisson statistics (and Cauchy-Bunyakovsky inequalities).

{\it The message of PCSFT in a nutshell is that (i) quantum systems may be mapped on classical stochastic systems even if they are capable of nontrivial quantum manifestations, and that (ii) this shows that the aforesaid phenomena should be regarded more classical than it is commonly believed.} 

Examples of mappings with the stated properties are well known: the $Q$-representation for linear bosonic systems, and the so-called positive-$P$ representation for nonlinear olistic nes. The $Q$-function of an electromagnetic field in a quantum state is positive, which does not preclude such field from showing violations of Bell inequalities in A. Aspect's experiments. More generally the {\it quantum tomographic approach}, see, e.g., V. I. Manko and collaborators \cite{M1}--\cite{Manko1}  provides a possibility to represent all quantum averages and correlations 
by using classical probability distributions. Although EPR correlations can be represented as correlations of classical random variables, as, e.g.,  in the quantum  tomographic approach and PCSFT, the violation of Bell's inequality is not surprising: some assumptions which had been used by Bell in the derivation of his ``theorem'' are violated.\footnote{For example, see I. V.  Volovich \cite{Volovich}  for analysis of the role of space-time dependence of correlations, see also monograph \cite{Volovich1}.}  In particular, PCSFT violates the spectral postulate: 
``prequantum variables'' (given by quadratic forms of classical fields) do not have the same range of values as their quantum 
counterparts. The assumption that observables take values $\pm1$ plays a crucial role in the original Bell's inequality for
correlations \cite{KHRCONT}. The CHSZ-inequality is based on a weaker assumption on the range of values of observables: they should take values
in the segment $[-1, +1].$ However, even the latter is violated in PCSFT. Hence, the main unsolved problem 
of PCSFT is creation of a novel measurement theory which will reproduce QM-measurement results from interaction of classical
fields. The first steps in this direction have already been done \cite{KH14}, and especially \cite{ArX}. 

We start this paper with a brief presentation of PCSFT.  This is  essentially theory of classical random fields, see 
\cite{KH1}--\cite{KH14}, \cite{KH1b}-- \cite{KH1b2}. 
However, to couple it with QM we have to rewrite this theory by using the language of operators and traces. Finally, in section \ref{LLKK}
we represent quantum correlations as correlations of {\it quadratic forms} 
of classical random signals. In section \ref{F1} we discuss the roles of phase in PCSFT and QM. By PCSFT formalism the wave function $\Psi$ of 
a composite quantum system $S=(S_1, S_2)$ determines the covariance operator $D_\Psi$ of the corresponding prequantum random field.
It is well known that in QM a pure state vector  is determined up to the phase factor. However, two vectors representing the same quantum pure state and having different phases (up to $2\pi)$ determine two different prequantum fields. And they produce the same correlations
for quadratic forms, i.e., reproduce the same quantum correlation. Then we consider, section \ref{Bos},  a class of classical random bi-signals (permutation invariant bi-signals) which reproduce bosonic quantum correlations. If a bi-signal is invariant under permutation of components
combined with the $\pi$-shift of the relative phase between components, section \ref{Fer}, 
then it reproduces fermionic correlations. In section \ref{bunch1} we study (anti-)bunching of spinless random fields interacting with beam splitters and in section \ref{bunch2} (anti-)bunching 
of spin 1/2 classical random fields is presented. Section \ref{SER} is devoted to representation of some quantum channels as classical signal channels, cf. \cite{W}, \cite{W1}.

The real physical state space of classical (as well as quantum) signals is the $L_2$-space. This space is infinite dimensional and
theory of random variables in this space (random fields) is  mathematically complicated. Therefore the reader can proceed
by considering finite dimensional Hilbert spaces and random variables valued in these spaces. In PCSFT it is convenient to operate 
with Gaussian random fields. In the finite dimensional case these are simply (complex valued) Gaussian random variables

The reader who is not so much interested in the problem of completeness of quantum mechanics and existence of hidden variables (and may be neither quantum foundations in general) can consider this paper 
as a contribution to classical probabilistic modeling of quantum phenomena, cf.  Ozhigov \cite{Ozhigov}--\cite{Ozhigov2}.

\section{Classical field representation of non-composite quantum systems}

{\bf Classical fields as hidden variables:}
Classical fields are  selected as hidden variables.
Mathematically these are functions $\phi: {\bf R}^3 \to {\bf C}$
(or more generally $\to {\bf C}^k)$ which are square integrable,
i.e., elements of the $L_2$-space.  

A random field (at the fixed instant of time)  is a function
$\phi(x, \omega),$  where $\omega$ is the random parameter. Thus,
for each $\omega_0,$ we obtain the classical field, $x\mapsto
\phi(x, \omega_0).$ 

By PCSFT each quantum system is a symbolic representation of 
a classical random field, a {\it prequantum field.} The state space of 
our prequantum model is the same as in the classical signal theory:
$H=L_2({\bf R}^3).$

By applying a linear functional $y$ to the
random vector $\phi$ we obtain the scalar random variable. In the
$L_2$-case we get a family of scalar random variables:
$
\omega \mapsto \xi_y(\omega) \equiv \int y(x) \overline{\phi(x,\omega)} dx, y \in L_2.
$
We recall that the covariance operator $D$ of a random field (with zero average)  $\phi \equiv  \phi(x,\omega)$ is defined by its bilinear form:
$
\langle Du, v\rangle = E \langle u, \phi \rangle  \langle \phi, v \rangle, u, v\in H.
$
Under the additional assumption that the prequantum random fields
are {\it Gaussian}, the covariance operator uniquely determines
the field.  We shall poceed under the {\it assumption that prequantum fields 
are Gaussian.}

We also suppose that that all {\it prequantum fields have zero average:}
$E\langle y, \phi \rangle= 0, y \in H,$
where $E$ denotes the classical {\it mathematical expectation}
(average, mean value). 

{\bf Covariance operator interpretation of wave function:}
In our model the wave function $\psi$ of the QM-formalism
encodes a prequantum random  field: $\phi\equiv \phi_\psi.$ The
QM-terminology, ``a quantum system in the state $\psi$'', is
translated into the PCSFT-terminology, a ``random field.'' In PCSFT
the {\it $\psi$-function determines the covariance operator of the
prequantum random field.} For simplicity, we consider the case of
a single, i.e., noncomposite, system, e.g., photon or electron.
In this situation  normalization (by dispersion) of the covariance operator $D$ of the prequantum field
is given by the orthogonal projector
on the vector $\psi$ (the density operator corresponding to this pure state)
\begin{equation}
\label{SEA6}
\rho_\psi= \psi \otimes \psi,
\end{equation}
i.e., $\rho_\psi u = \langle u, \psi\rangle \psi, \; u \in H.$ 
The covariance operator of the prequantum field is given by 
$
D= \sigma^2(\phi) \rho_\psi,
$
where 
$ \sigma^2(\phi)= E \Vert \phi \Vert^2 (\omega)= \rm{Tr} D $
is the dispersion of the prequantum random field $\phi$ which is distributed $N(0,D)$
(normally  with zero average and the covariance operator $D).$

To determine the covariance operator $D$ on the basis of the density operator $\rho_\psi,$ one has to find the scale of fluctuations of the prequantum field given by the dispersion $\sigma^2(\phi).$  

A prequantum random field $\phi(x,\omega)$ 
is not $L_2$-normalized. Its {\it intensity} (the $L_2$-norm)
\begin{equation}
\label{SEA3}
I(\phi(\omega))=  \Vert \phi \Vert^2 (\omega) \equiv \int_{{\bf R}^3} \vert \phi(x, \omega) \vert^2 dx
\end{equation}
fluctuates depending on the random parameter $\omega.$ We are interested in the average of this quantity
$\langle I \rangle \equiv E I(\phi).$
From the probabilistic viewpoint, it is nothing else than dispersion of the random field:
$\langle I \rangle= \sigma^2(\phi). $
Hence, to determine the scale of prequantum fluctuations, we have to find the mean intensity 
of this field. Since we are not able to monitor fluctuations of the prequantum fields corresponding to 
quantum systems, we cannot (at the moment) determine aforementioned scale. We proceed (as it is typical in 
probability theory) with normalized prequantum fields, i.e., we perform the transformation
$\phi \to \frac{\phi}{\sigma}.$

{\bf Quantum observables from quadratic forms of the prequantum
field:} In PCSFT quantum observables are represented by corresponding
quadratic forms of the prequantum field.  A self-adjoint operator
$\widehat{A}$ is considered as the symbolic representation of the
PCSFT-variable:
$\phi \mapsto f_A(\phi) = \langle \widehat{A} \phi, \phi \rangle.$
We remark that $f_A$ can be considered as a function on the phase space
of classical fields: $f_A\equiv f_A(q,p),$ where $\phi(x)=q(x)+ip(x),
q, p \in L_2({\bf R}^3; {\bf R}),$ the space of real valued fields.

Consider the quantum average
$\langle f_A \rangle_{{\rm{QM}}} =\langle \widehat{A} \psi, \psi \rangle.$
and the classical average
$\langle f_A \rangle_{{\rm{CL}}}= Ef_A(\phi)$
They coincide up to a scaling factor
$\langle f_A \rangle_{{\rm{CL}}}= \rm{Tr} D \; \langle f_A \rangle_{{\rm{QM}}}.$
For normalized prequantum fields, they simply coincide.

\section{Operator representation of the pure state of a composite system}

This section is a mathematical completion of the standard formalism of QM. 
The usage of coming new mathematical representation of basic expressions of 
QM will play an important role in constructing coupling between quantum correlations and correlations
of classical random fields representing composite quantum systems.

In this section we present the operator interpretation of pure states of a composite quantum system   $S=(S_1, S_2).$
In the canonical mathematical formalism of QM pure states are represented by vectors (normalized) 
belonging to the tensor product $H_1\otimes H_2$ of state spaces of subsystems $S_1$ and $S_2.$ In our approach
they are represented by linear operators acting from $H_2 \to H_1.$  

Let $\{e_j\}$ and $\{f_j\}$ be two orthonormal bases in $H_1$ and $H_2,$
respectively. Take $\Psi \in H_1 \otimes H_2:
\Psi = \sum_{ij} \Psi_{ij} e_i \otimes f_j, \Psi_{ij} \in {\bf C},
||\Psi||^2 = \sum_{ij} |\Psi_{ij}|^2 =1.$  For a vector $\phi \in H_2,$ we set
\begin{equation}
\label{AO}
\widehat \Psi \phi = \sum_{ij} \Psi_{ij} \langle\phi_2, \overline{f}_j \rangle e_i
\end{equation}
(the use of the combination $\overline{f}_j$ and  $e_i$ is important to get formulas which are consistent  with 
the vector representation of pure states of composite systems in the canonical mathematical framework of QM). 
Here $\widehat \Psi: H_2 \to H_1$ is the operator representation of the pure state 
$\Psi \in H_1 \otimes H_2$ of a composite quantum system $S=(S_1, S_2).$

Consider now very special, but at the same time very important case: the spaces of square integrable complex valued functions
$H_i=L_2({\bf R}^{n_i}), i=1,2.$ Here our definition gives the following
representation
\begin{equation}
\label{AOLK1}
\widehat \Psi \phi(x) = \int \Psi(x,y) \phi(y) d y
\end{equation}
Here we proceed formally and use the {\it real basis} $\{e_x\otimes f_y\},$ where
$e_x(t)= \delta(t-x)$ and $f_y(s)=\delta(s-y).$

{\bf Operation of the complex conjugation in the space of operators:}
Consider a linear operator  $\widehat{A}: H_2 \to H_1.$
 We define a ``complex conjugate operator'' $\widehat{\bar{A}}$ by its bilinear form:
\begin{equation}
\label{AOC}
\langle   \widehat{\bar{A}} u, v \rangle = \langle \bar{v}, \widehat{A} \bar{u}\rangle. 
\end{equation}
For a real basis, we have $\langle   \widehat{\bar{A}} e_i, e_j \rangle = 
\langle e_j, \widehat{A} e_i\rangle = \overline{\langle \widehat{A} e_i, e_j\rangle}.$ 
Hence, matrix elements of the complex conjugate operator $ \widehat{\bar{A}}$  are conjugate 
to matrix elements of the operator $\widehat{A}.$ (Operator is called real if $ \widehat{\bar{A}}=
 \bar{A};$ in a real basis matrix elements of a real operator are real numbers. For real operators,  coming consideration 
is essentially simpler. However, we cannot restrict our formalism to real operators. For example, the position operator and 
all Hamiltonians are real, but the momentum operator is not. Moreover, in general 
the operator $\widehat{\Psi}$  corresponding to a pure state $\Psi$ is not real.)
Definition (\ref{AOC}) implies
\begin{equation}
\label{AOC1}
 \langle \widehat{A}^* \bar{v},  \bar{u}\rangle= \langle   \widehat{\bar{A}} u, v \rangle 
\end{equation}
Since $\overline{\widehat{A}^*}= \widehat{\bar{A}}^*,$ we have
\begin{equation}
\label{AOC2}
 \langle \widehat{A}\bar{v},  \bar{u}\rangle= \langle   \widehat{\bar{A}}^* u, v \rangle.  
\end{equation}

Consider the quadratic form of the complex conjugate operator
$\widehat{\bar{A}}$ of a self-adjoint operator $\widehat{A}$
\begin{equation}
\label{AOC1N}
f_{\bar{A}}(\phi) =\langle \widehat{\bar{A}} \phi, \phi\rangle= \langle  \bar{\phi},  \widehat{A}\bar{\phi}\rangle=
f_{A}(\bar{\phi}).
\end{equation}
Consider the group $\{e, *\},$ where $* $ is the operation of
complex conjugation in a complex Hilbert space. It induces the
action in the space of real-valued functions on $W: f\to \bar{f},$
where $\bar{f}(\phi)= f(\bar{\phi}).$ (We hope that the symbol
$\bar{f}$ will not be misleading. Only real-valued functions are
under consideration. Thus it cannot be mixed with the operation of
complex conjugation in the range of values.)

{\bf The basic operator equality for self-adjoint operators:}
Let $\Psi\in H_1\otimes H_2.$ Then for any pair of self-adjoint (bounded) 
operators $\widehat{A}_j: H_j \to  H_j, j= 1,2, $
\begin{equation}
\label{01K}
\rm{Tr} \widehat{\Psi} \widehat{\bar{A}}_2 \widehat{\Psi}^* \widehat{A}_1=
\langle \widehat{A}_1 \otimes \widehat{A}_2 \rangle_\Psi \equiv
\langle \widehat{A}_1 \otimes \widehat{A}_2 \Psi, \Psi \rangle.
\end{equation}
The right-hand side of this equality gives the quantum average of the observable 
$\widehat{A}_1 \otimes \widehat{A}_2.$ It is based on the canonical vector interpretation of the
pure state $\Psi$ of a composite system $S=(S_1, S_2).$ The left-hand side provides 
the operator representation of this quantum average. We shall use this operator representation
in the process of transition from our prequantum model to QM. (We state again that in this 
section we just explore a new mathematical structure for QM, namely, the operator representation 
of pure states.)

{\bf Operator representation of reduced density operators:}
Consider the density operator  $\rho\equiv \rho_\Psi=\Psi \otimes \Psi$ and corresponding reduced density operators 
$\rho_i \equiv {\rm Tr}_{H_i} \rho_\Psi, i=1,2.$ The
following equalities hold:
\begin{equation}
\label{00} \rho_1 = \widehat \Psi \widehat\Psi^*, \; \bar{\rho}_2 = \widehat \Psi^* \widehat
\Psi.
\end{equation}
The appearance of the complex conjugation in the last equality plays an important role in consistent 
coupling of quantum and prequantum correlations.  We remark that, for any self-adjoint operator $\widehat{A},$
\begin{equation}
\label{00PT1}
\rm{Tr} \; \overline{\rho}\widehat{\bar{A}}= \rm{Tr} \; \rho \widehat{A}.
\end{equation}

\section{Classical random field representation of quantum correlations}
\label{LLKK}

Consider a composite quantum system $S=(S_1, S_2).$  Here $S_j$ has the state
space $H_j,$ a complex Hilbert space. Let $\phi_1 (\omega)$ and $\phi_2 (\omega)$
be two Gaussian random fields (with zero averages) in Hilbert spaces $H_1$ and $H_2,$ respectively.
Consider the Cartesian product of these Hilbert spaces, $H_1 \times H_2,$ and
the vector Gaussian random field
$\phi (\omega) = (\phi_1 (\omega), \phi_2 (\omega)) \in H_1 \times H_2.$
In the case under consideration its covariance operator has the block structure
given by \begin{equation}
\label{BSTR}
D = \left( \begin{array}{ll}
 D_{11} & D_{12}\\
D_{21} & D_{22}\\
 \end{array}
 \right ),
\end{equation}
where $D_{ii}: H_i \to H_i, D_{ij}: H_j \to H_i.$
The covariance operator is self-adjoint. Hence $D_{ii}^* = D_{ii},$ and
$D_{12}^* = D_{21}.$ The diagonal blocks $D_{ii}$ describe internal correlations in
subsystems. The block $D_{12}$ describes cross-correlations between subsystems. Here, by definition,
$$
\langle D_{ij} u_j, v_i \rangle = E \langle u_j, \phi_j (\omega)\rangle \langle \phi_i
(\omega), v_i\rangle , u_i \in H_i, v_j \in H_j.
$$ 
Set
$$
\langle f_{A_1}, f_{A_2}\rangle \equiv  E f_{A_1}  \bar{f}_{A_2}=
\int_{H_1\times H_2} f_{A_1}(\phi_1)  \bar{f}_{A_2} (\phi_2) d \mu
(\phi_1, \phi_2)
$$
$$
=\int_{H_1\times H_2} f_{A_1}(\phi_1)  f_{A_2}(\overline{\phi}_2)
d \mu (\phi_1, \phi_2),
$$
where $\mu$ is the probability distribution of the vector random field $\phi.$ 
Set also
$$
{\rm cov} \; (f_{A_1}, f_{A_2})= \langle f_{A_1}, f_{A_2}\rangle - \langle f_{A_1}\rangle \langle \bar{f}_{A_2}\rangle.
$$
It is possible to prove (by calculating Gaussian integrals)
that, for any Gaussian
random field (with zero average) $\phi(\omega)$ in $H_1\times H_2$ with the covariance matrix $D$
such that the off-diagonal block has the form
\begin{equation}
 \label{HHUURR}
D_{12}= \widehat{\Psi}
\end{equation}
the following equality takes place:
 \begin{equation}
 \label{00T5}
{\rm cov} \; (f_{A_1}, f_{A_2}) = (\widehat{A}_1 \otimes \widehat{A}_2 \Psi, \Psi) \equiv
\langle \widehat{A}_1 \otimes \widehat{A}_2 \rangle_\Psi.
\end{equation}
This equality establishes a coupling between the quantum and classical
correlations.

{\bf Diagonal blocks of the covariance operator of a prequantum random field:}
The operators $D_{ii}$ are responsible for averages of functionals depending
only on one of the components of the vector random field $\phi (\omega).$ In
particular, $E f_{A_1} (\phi_1) (\omega)) = {\rm Tr} D_{11} \widehat{A}_1$
and $E \bar{f}_{A_2} (\phi_2) (\omega)) = {\rm Tr} D_{22} \widehat{\bar{A}}_2.$
We will construct a random field such that these ``marginal averages'' will
match those given by QM. For the latter, we have
$
\langle \widehat{A}_1 \rangle_\Psi = (\widehat{A}_1 \otimes I\Psi,
\Psi) = {\rm Tr} \rho_1 \widehat{A}_1, \langle
\widehat{A}_2 \rangle_\Psi = (I \otimes \widehat{A}_2 \Psi,
\Psi)={\rm Tr} \rho_2 \widehat{A}_2,
$
where $I$ denotes the unit operator.  By by the first equality in (\ref{00}) the
first average can be written as
$$
\langle \widehat{A}_1 \rangle_\Psi={\rm Tr} (\widehat{\Psi} \widehat{\Psi}^*) \widehat{A}_1,
$$
by (\ref{00PT1}) the second average can be represented as
$
\langle \widehat{A}_2 \rangle_\Psi = {\rm Tr} \overline{\rho}_2 \widehat{\bar{A}}_2
$
and, finally, by the second equality in (\ref{00}) we obtain:
$$
\langle \widehat{A}_2 \rangle_\Psi
  = {\rm Tr} (\widehat{\Psi}^* \widehat{\Psi}) \widehat{\bar{A}}_2,
$$
Thus it would be natural to take
$D = \left( \begin{array}{ll}
 \widehat{\Psi} \widehat{\Psi}^* & \; \; \widehat{\Psi} \\
 \; \; \widehat{\Psi}^* & \widehat{\Psi}^*\widehat{\Psi}
 \end{array}
 \right ).
$
Its off-diagonal block reproduces correct quantum correlations between
systems $S_1$ and $S_2$, and its diagonal blocks produce correct quantum
averages for system $S_1$ and system $S_2.$
However, in general (i.e., for an arbitrary pure state $\Psi)$ this operator is
not positively defined.  Let us consider a modification which will
be positively defined and such that quantum and classical averages will be coupled
by a simple rule. Thus from quantum averages one can easily find classical averages
and vice versa.  For any normalized vector $\Psi \in H_1 \otimes H_2$ and
sufficiently large $ \epsilon > 0,$ the operator
\begin{equation}
\label{YYYO}
D\equiv D_\Psi = \left( \begin{array}{ll}
 (\widehat{\Psi} \widehat{\Psi}^* + \epsilon I) & \; \; \; \; \; \; \; \widehat{\Psi}\\
\; \;  \; \; \; \; \; \widehat{\Psi}^* &
(\widehat{\Psi}^*\widehat{\Psi} +  \epsilon I)
 \end{array}
 \right )
\end{equation}
{\it is positively defined,} see \cite{KH1b}-- \cite{KH1b2}. The modification of the diagonal terms can be interpreted as taking into account
the presence of a background field; a random field of the white noise type: Gaussian field with zero average and the covariance operator
$D_{\rm{background}}= \epsilon I.$ It is impossible to construct a prequantum random field matching quantum correlations without 
taking into account the presence of the background field (``fluctuations of vacuum''). This is the crucial point of of our theory.
Mathematically the situation is even more tricky. It is impossible to distill the  ``intrinsic field'' of a quantum system, e.g. an electron, from the random background. Such an ``intrinsic electronic field'' does not exist, it is meaningful only in the combination with the background field. This is the ontic situation, i.e., as it is in nature (without any relation to our measurements).
At the same time it is clear that in measurement theory one has to eliminate the contribution of vacuum fluctuations. This was done in the canonical QM. (Following Bohr, we consider QM as theory of measurements.) We shall discuss this crucial point in more detail. 

For the Gaussian measure with the covariance operator (\ref{YYYO}), we have
\begin{equation}
\label{YY1} \langle \widehat{A}_1 \rangle_\Psi = E f_{A_1} (\phi_1
(\omega)) - \epsilon {\rm Tr} \widehat{A}_1,
 \end{equation}
\begin{equation}
\label{YY2} \langle \widehat{A}_2 \rangle_\Psi = E f_{A_2} (\phi_2
(\omega)) - \epsilon {\rm Tr} \widehat{A}_2.
\end{equation}
These relations for averages and relation (\ref{00T5}) for the correlation
provide coupling between PCSFT and QM. Quantum statistical
quantities can be obtained from corresponding quantities for
classical random field. One may say that
{\it ``irreducible quantum randomness" is
reduced to randomness of classical prequantum fields,} cf. von Neumann \cite{VN}. However, the situation is more complicated.

The equalities (\ref{YY1}), (\ref{YY2}) imply that quantum
averages are obtained as the shift-type renormalizations of
averages with respect to classical random fields. The shift
corresponds to elimination of the contribution of the background
field. Thus quantum averages are not simply classical averages.
Therefore, in some sense von Neumann was right \cite{VN}. Quantum averages
are not represented in the form
\begin{equation}
\label{YY3} \langle \widehat{A}_2 \rangle_\Psi = E g_A (\phi(\omega)),
\end{equation}
where $\phi_\Psi \equiv \phi(\omega)$ is a prequantum random field corresponding to a quantum state $\Psi$ and $g_A(\phi)$ a functional
of the prequantum field. By ignoring mathematical difficulties induced by the infinite dimension we can write the equalities
(\ref{YY1}), (\ref{YY2}) as
\begin{equation}
\label{YY1a} \langle \widehat{A}_1 \rangle_\Psi = E f_{A_1} (\phi_1
(\omega)) - \epsilon E f_{A_1} (\eta(\omega)),
 \end{equation}
\begin{equation}
\label{YY2a} \langle \widehat{A}_2 \rangle_\Psi = E f_{A_2} (\phi_2
(\omega)) - \epsilon E f_{A_2} (\eta(\omega)),
\end{equation}
where $\eta \sim N(0, I)$ is white noise: the Gaussian random variable with zero mean value and the unit covariance matrix. Hence, although we do not have
the representation (\ref{YY3}), the representation  (\ref{YY1a}), (\ref{YY2a}) can be interpreted as a reduction of quantum randomness to
classical ensemble randomness. Therefore, in some sense Einstein was right as well \cite{EI}. In PCSFT framework the positions of von Neumann and
Einstein have been peacefully unified. However, I am not sure that either von Neumann or Einstein would be happy with such a peaceful
agreement between the Copenhagen and classical camps. Von Neumann was definitely sure that the mystery of quantum randomness is not
reduced to the shift-type transformation    (\ref{YY1a}), (\ref{YY2a}). And Einstein was not interested in the reduction of quantum effects
to a random background. In any event he did not support Planck's attempt to explain spontaneous emission by a random disturbance effect
induced by the background field.

At the same time the presentation (\ref{YY1a}), (\ref{YY2a}) of quantum averages matches quantum experimental science which uses
the procedure of {\it calibration of detectors.} \index{calibration of detector} Subtraction of the contribution of the random background is a theoretical counterpart
of the calibration procedure.

\section{Role of phase of wave function in quantum and prequantum theories}
\label{F1}

 In the canonical quantum formalism a pure state vector
is determined up to a  phase factor. Two normalized vectors
$\Psi_1, \Psi_2 \in H$ determine the same pure quantum state iff
\begin{equation}
\label{BOS1}
\Psi_2 = e^{i \theta} \Psi_1.
\end{equation}
For a single particle system, the same thing happens in PCSFT. The
covariance operator $D_\Psi\equiv \rho_\Psi = \vert \Psi \rangle
\langle \Psi\vert$ of a prequantum random field $\phi(\omega)
\equiv \phi_\Psi(\omega)$ does not depend on the phase factor;
$D_{\Psi_1}= D_{\Psi_2}$ for $\Psi_1, \Psi_2$ coupled by
(\ref{BOS1}). Hence, if we restrict consideration to Gaussian
prequantum fields, such $\Psi_1$ and  $\Psi_2$ determine the same
field. However, for composite systems the roles of the phase
factor are totally different. In QM the transition to composite
systems changes nothing; in PCSFT two vectors coupled by
(\ref{BOS1}) determine two different random bi-signals (even in
the Gaussian  case); they  have covariance operators with
different off-diagonal blocks. We remark that if two vectors from $H_1\otimes H_2$ are coupled 
by the relation (\ref{BOS1}), then the corresponding operators are coupled by the similar relation: 
\begin{equation}
\label{BOS1k}
\widehat{\Psi}_2 = e^{i \theta} \widehat{\Psi}_1.
\end{equation}
Hence, we have
\begin{equation}
\label{YYYOk}
D_{\Psi_1} = \left( \begin{array}{ll}
 (\widehat{\Psi_1} \widehat{\Psi_1}^* + \epsilon I) & \; \; \; \; \; \; \; \widehat{\Psi_1}\\
\; \;  \; \; \; \; \; \widehat{\Psi_1}^* &
(\widehat{\Psi_1}^*\widehat{\Psi_1} +  \epsilon I)
 \end{array}
 \right ),
\end{equation}
\begin{equation}
\label{BOS1k2}
D_{\Psi_2} = \left( \begin{array}{ll}
 (\widehat{\Psi_1} \widehat{\Psi_1}^* + \epsilon I) & \; \; \; \; \; \; \; e^{i \theta} \widehat{\Psi_1}\\
\; \;  \; \; \; \; \; e^{-i \theta} \widehat{\Psi_1}^* &
(\widehat{\Psi_1}^*\widehat{\Psi_1} +  \epsilon I)
 \end{array}
 \right ). 
\end{equation}
We remark that the diagonal blocks do not depend on the phase.  Thus in PCSFT the phase of a normalized vector representing
a pure quantum state (of a composite system) does not play any role for internal correlations inside each component of a bi-signal
$\phi=(\phi_1, \phi_2).$ It is important only for correlations between different components.

Let $\Psi \in H_1 \otimes H_2, \Vert \Psi \Vert^2=1.$   Consider a random bi-signal $\phi(\omega)\equiv \phi_\Psi(\omega)$ with the covariance operator
$D\equiv D_\Psi.$  Take a pair of phases $\gamma=(\gamma_1, \gamma_2), \gamma_j \in [0, 2\pi).$ Consider
the phase transformation in the {\it Cartesian  product} of Hilbert spaces: $\Phi_\gamma \phi= (e^{i\gamma_1} \phi_1,
e^{i\gamma_2} \phi_2)$. We now apply this transformation to the bi-signal $\phi(\omega);$ the output signal is give by
\begin{equation}
\label{BOS2}
\phi_{\gamma}(\omega)= \Phi_\gamma \phi(\omega).
\end{equation}
The covariance operator of the output $\phi_{ \gamma}$ can be
easily found; we are interested in its off-diagonal block (the
diagonal blocks are preserved):
$$
\langle D^{\phi_{\gamma}}_{12} y_2, y_1\rangle= E \langle y_2,
e^{i\gamma_2} \phi_2 \rangle \langle e^{i\gamma_1}  \phi_1, y_1
\rangle = \langle D_{12} e^{-i\gamma_2} y_2,
e^{-i\gamma_1} y_1\rangle.
$$
Hence,
\begin{equation}
\label{BOS3}
D^{\phi_{\gamma}}_{12} = e^{i(\gamma_1- \gamma_2)} D_{12}.
\end{equation}
Since $D_{12}= \widehat{\Psi},$ the output signal $\phi_{\gamma}$
corresponds to the vector
\begin{equation}
\label{BOS4}
\Psi_\gamma = e^{i(\gamma_1- \gamma_2)} \Psi.
\end{equation}
Thus the phase transformation of a bi-signal does not change its probability distribution iff
the relative phase $\gamma_1 - \gamma_2= 2\pi k.$

This result can be used to illustrate the problem of the phase
choice in PCSFT. Let $\phi$ be a bi-signal corresponding to a
vector  $\Psi_1.$ Then any bi-signal $\phi_{\gamma}$ can be
considered as a prequantum representation of the same quantum
state. Thus each pure quantum state represents, in fact, the two
parameter family $\phi_{\gamma}$ of prequantum random fields
(coupled by phase transformations). However, QM is not aware about
this, since at the level of quantum observables the phase
dependence of prequantum random signals is not visible.

In PCSFT quantum observables are represented by quadratic forms of prequantum signals.
For a pair of self-adjoint operators $\widehat{A}_k: H_k \to  H_k,$ we have $f_{A_k}(\phi_{\gamma,k})=
f_{A_k}(\phi_{k}), k=1,2,$ where $\phi_{\gamma}=  (\phi_{\gamma,1}, \phi_{\gamma, 2}).$ Hence,
$$ Ef_{A_1}(\phi_{\gamma,1}) f_{A_1}(\phi_{\gamma,2}) =E f_{A_1}(\phi_{1})f_{A_2}(\phi_{2}).$$ Thus at the level
of correlations of quadratic forms (and, hence, quantum observables)  the phase dependence of signals is not visible.

\section{Transformation of the covariance operator under permutation of signal's components}

Consider the permutation operator  in the Cartesian product $H \times H:$ $$\sigma (\phi_1, \phi_2)=
(\phi_2, \phi_1).$$ We can define the permutation transformation of a prequantum random field,
$\phi^\sigma(\omega)= (\phi_2(\omega), \phi_1(\omega)).$

\subsection{Bosonic prequantum random signals}
\label{Bos}

 It were natural (at least for Gaussian signals) to call  a
random field {\it permutation symmetric}, or simply symmetric, if
random fields $\phi$ and $\phi^\sigma$ have the same covariance
operator. However, the definition of covariance involves the
operation of complex conjugation.  To match this (mathematical)
definition of covariance, we have to complete the permutation
transformation by the conjugation transformation, $* \phi=
\bar{\phi}.$ Set $\sigma_*= * \sigma.$ For a random field
$\phi(\omega),$ consider its transformation
\begin{equation}
\label{BOS5}
\phi^{\sigma_*}(\omega)= \sigma_* \phi(\omega) = (\bar{\phi}_2, \bar{\phi}_1).
\end{equation}

We call components of a random bi-signal $\phi$ {\it permutation
symmetric} or simply call $\phi$ a {\it symmetric bi-signal} if
$\phi$ and $\phi^{\sigma_*}$ have the same covariance operator.
(In the Gaussian case this means  ``the same probability
distribution''.) Denote covariance operators of aforementioned
bi-signals by symbols $D^{\phi}$ and $D^{\phi^{\sigma_*}},$
respectively. Thus $\phi$ is symmetric iff
\begin{equation}
\label{BOS5a} D^{\phi} = D^{\phi^{\sigma_*}}.
\end{equation}
Since the diagonal blocks of these operators coincide, see section \ref{OCOM} for details,  this
condition is reduced to the coincidence of the of-diagonal blocks:
\begin{equation}
\label{BOS5b} D^{\phi}_{12} = D^{\phi^{\sigma_*}}_{12}.
\end{equation}
We reformulate this condition in terms of quantum states. Let a
bi-signal $\phi$ correspond to   the normalized vector $\Psi,$ so
$\phi \equiv \phi_\Psi.$ We find the covariance bilinear form of
$\phi^{\sigma_*}:$
\begin{equation}
\label{BOS6}
b_{\sigma_*}(y_2, y_1)= E \langle y_2, \bar{\phi}_1\rangle \langle \bar{\phi}_2, y_1\rangle=
\langle D_{12} \bar{y}_1, \bar{y}_2 \rangle= \langle \widehat{\Psi} \bar{y}_1, \bar{y}_2 \rangle.
\end{equation}
A bi-signal $\phi\equiv \phi_\Psi$ corresponding to a pure quantum
state $\Psi$ is symmetric   if
\begin{equation}
\label{BOS7} \langle \widehat{\Psi} \bar{y}_1, \bar{y}_2 \rangle =
\langle \widehat{\Psi} y_2, y_1 \rangle.
\end{equation}
Take a real orthonormal basis $\{e_k\}$ in $H,$ i.e., $\bar{e}_k=
e_k.$ Then condition (\ref{BOS7}) is equivalent to the following
equality for the matrix elements of the operator $\widehat{\Psi}:$
\begin{equation}
\label{BOS8} \Psi_{ij}= \Psi_{ji}.
\end{equation}
If all matrix elements are real, then this operator is
self-adjoint; in the general case this is a more complex
restriction to the operator $\widehat{\Psi},$ see section \ref{OCOM}. 

In fact, we did not
get that the permutation symmetry is equivalent to the
self-adjointness of  a pure quantum state in the operator
representation, because complex structures in the tensor product
$H_1\otimes H_2$ and the space of operators ${\cal L}(H_2, H_1)$
do not match each other.  An operator form of the equality (\ref{BOS7}) will 
be presented in  section \ref{OCOM}.

In a real basis, the matrix elements of the
operator $\widehat{\Psi}$ coincide with the coordinates of the
vector $\Psi \in H_1\otimes H_2, \Psi= \sum_{ij} \Psi_{ij} e_i
\otimes e_j,$ see (\ref{AO}). Thus in this case (\ref{BOS8}) can be
interpreted as the equality of coordinates of the quantum state
vector.

Let $H=L_2({\bf R}^3).$ Here $\Psi= \Psi(x_1, x_2)$  is a square integrable function on
${\bf R}^6.$
Taking in (\ref{BOS8}) $y_1(u_1) = \delta (x_1 -   u_1), y_2(u_2) = \delta (x_2 -   u_2)$
we obtain
\begin{equation}
\label{BOS9} \Psi(x_1, x_2) =\Psi(x_2, x_1).
\end{equation}
Thus a prequantum random signal (with  one-dimensional projection
as the covariance  operator) is symmetric if it corresponds to a
symmetric wave function. The corresponding quantum system
``consists of two bosons.''  Therefore one may call permutation
symmetric prequantum fields {\it bosonic fields.}

The following example can be used to clarify the meaning of the
equality (\ref{BOS7}). Consider a pure quantum state of the form
\begin{equation}
\label{BOS10} \Psi= c_1\Psi_1\otimes \Psi_2 + c_2\Psi_2\otimes
\Psi_1,
\end{equation}
 where $\Psi_1, \Psi_2 \in H,  c_1, c_2 \in {\bf C},
\vert c_1\vert^2 + \vert c_2\vert^2= 1.$ Then
$$
\widehat{\Psi} y_2= c_1 \langle y_2, \bar{\Psi}_2\rangle \Psi_1 + c_2 \langle y_2, \bar{\Psi}_1\rangle \Psi_2;
$$
$$
\langle\widehat{\Psi} y_2, y_1 \rangle = c_1 \langle y_2, \bar{\Psi}_2\rangle\langle \Psi_1, y_1 \rangle
+ c_2 \langle y_2, \bar{\Psi}_1\rangle \langle\Psi_2, y_1\rangle;
$$
$$
\langle\widehat{\Psi} \bar{y}_1, \bar{y}_2 \rangle =
c_1 \langle \bar{y}_1, \bar{\Psi}_2\rangle\langle \Psi_1, \bar{y}_2 \rangle
+ c_2 \langle \bar{y}_1, \bar{\Psi}_1\rangle \langle\Psi_2, \bar{y}_2\rangle
$$
$$
= c_1 \langle  \Psi_2, y_1 \rangle\langle y_2, \bar{\Psi}_1
\rangle + c_2 \langle \Psi_1, y_1, \rangle \langle y_2,
\bar{\Psi}_2\rangle.
$$
Comparing these expressions we obtain the equality $c_1=c_2=1/\sqrt{2},$ i.e.
\begin{equation}
\label{BOS11} \widehat{\Psi} =  \frac{1}{\sqrt{2}} (\Psi_1\otimes
\Psi_2 + \Psi_2\otimes \Psi_1).
\end{equation}

\subsection{Fermionic prequantum random signals}
\label{Fer}

 We call components of a random bi-signal $\phi$ {\it
permutation anti-symmetric} or simply call $\phi$ an {\it
anti-symmetric bi-signal} if the covariance operators of $\phi$
and $\phi^{\sigma_*}$ have the same diagonal blocks and the
off-diagonal blocks have different signs:
\begin{equation}
\label{BOS5b1} D^{\phi^{\sigma_*}}_{12} =-  D^{\phi}_{12}
\end{equation}
By using the language of QM this condition can be rewritten as
\begin{equation}
\label{BOS71} \langle \widehat{\Psi} \bar{y}_1, \bar{y}_2 \rangle
= - \langle \widehat{\Psi} y_2, y_1 \rangle.
\end{equation}
For a real basis in $H,$ this condition is equivalent to the
following relation for the matrix elements of the operator
$\widehat{\Psi}$ (coordinates of the state vector $\Psi):$
\begin{equation}
\label{BOS8a} \Psi_{ij}= - \Psi_{ji}.
\end{equation}
If all matrix elements are real this is equivalent to
skew-symmetry of the operator  $\widehat{\Psi}.$

Take now $H=L_2({\bf R}^3);$ by proceedings similar to section
\ref{Bos} we obtain that the wave function $\Psi(x_1, x_2)$ of the
anti-symmetric bi-signal $\phi$ is anti-symmetric:
\begin{equation}
\label{BOS91} \Psi(x_1, x_2) =- \Psi(x_2, x_1).
\end{equation}
The corresponding quantum system ``consists of two fermions.''
Therefore one may call permutation symmetric prequantum fields
{\it fermionic fields.}

Consider a pure quantum state of the form (\ref {BOS10}). The
corresponding random bi-signal $\phi$ is anti-symmetric iff
$c_1=-c_2,$ i.e.,
\begin{equation}
\label{BOS111} \widehat{\Psi} =  \frac{1}{\sqrt{2}} (\Psi_1\otimes
\Psi_2 - \Psi_2\otimes \Psi_1).
\end{equation}
We now modify the transformation $\sigma_*$ by setting
\begin{equation} \label{BOS111q} \phi^{\sigma_{*-}}\equiv
\sigma_{*-} \phi = (-\bar{\phi}_2, \bar{\phi}_1).
\end{equation}
A bi-signal is anti-symmetric iff the signals $\phi$ and
$\phi^{\sigma_{*-}}$ have the same covariance operators:
\begin{equation}
\label{BOS5a1q} D^{\phi} = D^{\phi^{\sigma_{*-}}}.
\end{equation}
This is equivalent to the coincidence of the off-diagonal blocks
\begin{equation}
\label{BOS5b1qd} D^{\phi}_{12}= D^{\phi^{\sigma_{*-}}}_{12}.
\end{equation}
Thus a prequantum random field is fermionic if its covariance
operator is not changed after permutation of components and the
phase change $\Delta \gamma = \pi.$ By taking into account results
of section \ref{F1} we see that, in fact, it is important only the
change of  the relative phase of signal's components. Instead of
the transformation (\ref{BOS111q}), we can consider the
transformation ($\gamma= (\gamma_1, \gamma_2)):$
\begin{equation}
\label{BOS111qx} \phi^{\sigma_{*\gamma}}\equiv
\sigma_{*} \Phi_\gamma \phi = ( e^{-i\gamma_2} \bar{\phi}_2,
e^{-i\gamma_1} \bar{\phi}_1)
\end{equation}
with $\Delta \gamma = \gamma_1- \gamma_2 =  \pi.$

PCSFT is a formal classical field formalism for quantum phenomena.
It cannot explain why one class of signals exhibits symmetry of
its components and another class anti-symmetry. (We remark that
neither QM can explain why some particles are bosons and other are
ferimions.) In QM statistics of Bose and Fermi are often coupled
to {\it indistinguishability} of quantum systems. Consider now a
bosonic bi-signal; permutation symmetry of its components does
not imply indistinguishability. In principle, by improving
technology we might hope to be able to monitor components of a
signal and, hence, to distinguish them (in the same way as
classical particles). Permutation symmetry implies only that the covariance operator
is invariant under the exchange of signal's components (or in the Gaussian case the
probability distribution is symmetric). Permutation anti-symmetry implies that the covariance operator
is invariant under the exchange of signal's components combined with the  $\pi$-shift of the relative phase of
the components.

We can also consider prequantum random signals of the {\it anyonic}
type. Consider the signal transformation (\ref{BOS111qx}) with
$\Delta \gamma = \theta,$ where $\theta\in [0, 2 \pi)$ is fixed.
We call $\phi$ an anyonic bi-signal with the $\theta$-symmetry if
the bi-signals $\phi$ and $\phi^{\sigma_{*\gamma_1\gamma_2}}$ have the
same covariance operator (in the Gaussian case the same
probability distribution) or (equivalently) the bi-signals $\phi$
and $\phi^{\sigma_{*}}$ have the same diagonal blocks and the
off-diagonal blocks are coupled by
\begin{equation}
\label{BOS5b1qd} D^{\phi}_{12}= e^{i \theta}
D^{\phi^{\sigma_{*}}}_{12}.
\end{equation}
PCSFT-formalism does not imply that anyonic
fields for $\theta \not= \pi k$ do not exist.

\section{Operator representation of conditions of permutation symmetry and anti-symmetry}
\label{OCOM}

By using notion of complex conjugation in the space of operators, see (\ref{AOC}),
the equality (\ref{BOS6}) can be written as 
\begin{equation}
\label{yy}
D^{\phi^{\sigma_*}}_{12}= \widehat{\bar{\Psi}}^*,
\end{equation}
and the equality (\ref{BOS7}) as
\begin{equation}
\label{yy}
\widehat{\Psi} =
\widehat{\bar{\Psi}}^*
\end{equation}
or 
\begin{equation}
\label{yy1}
\widehat{\bar{\Psi}} =
\widehat{\Psi}^*.
\end{equation}
This is the operator representation of the condition of permutation symmetry. 

We now show (as was promised in section \ref{Bos})
that, for a bosonic bi-signal, the diagonal blocks  of its covariance operator
are not changed by the permutation transformation (\ref{BOS5}). We have
\begin{equation}
\label{BOS6j}
 \langle D^{\phi_{\sigma_*}}_{11} y_1, y_1\rangle= E \langle y_1, \bar{\phi}_2\rangle \langle \bar{\phi}_2, y_1\rangle=
\langle D_{22} \bar{y}_1, \bar{y}_1 \rangle =\langle \bar{D}^{\phi}_{22} y_1, y_1 \rangle .
\end{equation}
Hence, the equality $D^{\phi^{\sigma_*}}_{11}= D^{\phi}_{11}$ can be written as 
$D^{\phi}_{11} = \bar{D}^{\phi}_{22}.$ We have $D^{\phi}_{11}= \widehat{\Psi} \widehat{\Psi}^*$ and 
$\bar{D}^{\phi}_{22}= \widehat{\bar{\Psi}}^* \widehat{\bar\Psi}.$ By taking into account  (\ref{yy1})
we obtain $\bar{D}^{\phi}_{22}=\widehat{\Psi} \widehat{\Psi}^*.$ 

In the same way in the fermionic case the condition (\ref{BOS71}) can be written as 
\begin{equation}
\label{yy}
\widehat{\Psi} =
- \widehat{\bar{\Psi}}^*
\end{equation}
This is the operator representation of the condition of permutation anti-symmetry. It is clear 
that under this condition the diagonal blocks of the covariance operator are preserved (as in the case of 
the permutation symmetry).  

We state again, see section \ref{F1}, that any pure quantum state determines the two parametric family of 
prequantum (Gaussian) random fields $\phi_\gamma = \Phi_\gamma \phi, \gamma= (\gamma_1, \gamma_2),$ see (\ref{BOS2}),
where $\phi\equiv \phi_\Psi$ is the random field with the covariance operator $D_\Psi.$ We show that if one representative 
of this family of random fields is permutation symmetric (bosonic), then all random fields $\phi_\gamma$ are bosonic as well.
Suppose that e.g. $\phi\equiv \phi_\Psi$ is symmetric, i.e., (\ref{yy}) holds. By (\ref{BOS4}) the random field $\phi_\gamma$ corresponds
to the vector $\Psi_\gamma= e^{i\theta} \Psi,$ where $\theta= \gamma_1 - \gamma_2.$ We remark that, for any operator 
$\widehat{C}$ and its phase shift $\widehat{C}_\theta= e^{i\theta}\widehat{C},$ the following equality takes place:
$\widehat{\bar{C}}_\theta= e^{-i \theta} \widehat{\bar{C}}.$ We also know that $\widehat{C}_\theta^* = e^{-i \theta} \widehat{C}^*.$
Hence, if (\ref{yy}) holds, then $\widehat{\bar{\Psi}}_\gamma^* =\widehat{\Psi}_\gamma.$ Thus the random field $\phi_\gamma$ is symmetric.
Anti-symmetric random fields have the same property. Thus each bosonic (fermionic) quantum state determines a family of bosonic (fermionic)
prequantum random fields. These cannot be  distinguished on the level of the QM-model.

\section{Classical channel representation of quantum channels}
\label{SER}

As was found by Ohya and Watanabe \cite{W}, classical Gaussian signal channels
can be represented as quantum channels (this fact was used in the classical signal theory).
In \cite{W1} it was shown that some important quantum channels can be represented as classical channels.
(It is not clear whether any quantum channel can be represented as a classical channel.) In this paper
we are interested in the classical representation of quantum channels corresponding to unitary transformations.

{\bf Classical random field viewpoint to Schr\"odinger equation:} 
Before to go to the PCSFT-dynamics, we consider the Schr\"odinger equation in the standard
QM-formalism:
\begin{equation}
\label{SE}
i h \frac{\partial \Psi}{\partial t}(t)= \widehat{\cal H} \Psi (t),
\end{equation}
\begin{equation}
\label{SE1}
\Psi(t_0)=\Psi_0,
\end{equation}
where $\widehat{\cal H}$ is Hamiltonian.

We recall that a time dependent random field $\phi(t, x, \omega)$  is called the {\it stochastic
process}  (with the state space $H)$. The dynamics of the prequantum random field
 is described by the simplest stochastic process which is given by {\it deterministic dynamics with
 random initial conditions.}

In PCSFT the Schr\"odinger equation, but with the random initial condition, describes
the dynamics of the prequantum random field, i.e., the prequantum stochastic process can be obtained
from ``Cartesian product representation'' of the Schr\"odinger equation:
\begin{equation}
\label{SE2}
i h \frac{\partial \phi}{\partial t}(t, \omega)= \widehat{\cal H}_{\rm{PCSFT}}\phi (t, \omega),
\end{equation}
\begin{equation}
\label{SE3}
\phi(t_0, \omega)=\phi_0(\omega).
\end{equation}
Here $\widehat{\cal H}_{\rm PCSFT}: H_1\times H_2 \to H_1\times H_2$ is the PCSFT-generator of dynamics
corresponding to the quantum generator $\widehat{\cal H}: H_1 \otimes H_2 \to H_1 \otimes H_2.$
The initial random field $\phi_0(\omega)= (\phi_{01}(\omega), \phi_{02}(\omega))$ is determined by the quantum pure state $\Psi_0.$ This is a field with zero mean value and the covariance operator $D_{\Psi_0},$  see (\ref{YYYO}).

For quantum Hamiltonian without interaction,
\begin{equation}
\label{SE2q}
\widehat{\cal H}= \widehat{\cal H}_1\otimes I + I \otimes \widehat{\cal H}_2,
\end{equation}
 its PCSFT-image can be easily found,
$\widehat{\cal H}_{\rm PCSFT}= \widehat{\cal H}_1 \times \widehat{\cal H}_2,$ i.e., dynamics
(\ref{SE3}) can be written in the form of the system of equations for components of a random field:
\begin{equation}
\label{SE2j}
i h \frac{\partial \phi_1}{\partial t}(t, \omega)= \widehat{\cal H}_1 \phi_1 (t, \omega),
\end{equation}
\begin{equation}
\label{SE2ji}
i h \frac{\partial \phi_2}{\partial t}(t, \omega)= \widehat{\cal H}_2 \phi_2 (t, \omega),
\end{equation}
\begin{equation}
\label{SE3j}
\phi_1(t_0, \omega)=\phi_{01}(\omega), \phi_2(t_0, \omega)=\phi_{02}(\omega).
\end{equation}
Set $U_{jt}=\exp\{-it \widehat{\cal H}_j/h\}, j=1,2,$ and
\begin{equation}
\label{SE3jmu}
V_t = U_{1t} \times U_{2t}.
\end{equation}
Then $\phi_t= V_t \phi_0.$
It is easy to see that the off-diagonal block of the random field
\begin{equation}
\label{SE2ji1}
\phi_t(\omega) = V_t \phi_0(\omega)
\end{equation}
is given by the operator $\widehat{\Psi_t},$
where $\Psi_t= U_t \Psi_0$ and $U_t= U_{1t} \otimes U_{2t}.$  In this way the equation (\ref{SE2})
for a classical random process matches the Schr\"odinger equation of QM.

We remark that, in spite of the representation (\ref{SE3jmu}), dynamics (\ref{SE2j}),(\ref{SE2ji}), (\ref{SE3j}) (as the dynamics of random fields) cannot be split into two independent dynamics, for components of the random field. In general the initial fields $\phi_{01}(\omega)$ and $\phi_{02}(\omega)$ are not independent.

In general, any quantum channel given by the unitary operator $U=U_1\otimes U_2$ can be represented as
the unitary transformation of a classical random signal:
\begin{equation}
\label{SE2ji2}
\phi_{\rm{out}}(\omega) = (U_1 \phi_{\rm{in}, 1}(\omega), U_2 \phi_{\rm{in}, 2}(\omega)).
\end{equation}
For the input signal corresponding to a pure quantum state
$\Psi_{\rm{in}},$ the off-diagonal block of $\phi_{\rm{out}}$ is
given by
\begin{equation}
\label{SE2ji3}
\Psi_{\rm{out}}= U \Psi_{\rm{in}}.
\end{equation}

\section{Bunching and anti-bunching  in quantum theory}

Let us consider a following thought experiment. Two identical quantum particles are
incident upon a 50\% -- 50\% particle beam splitter (BS) and we measure the output by two particle
counters).

To analyze this problem, we label the two particles as particle 1 and particle 2 based on
the click event in the detectors. One possible input state is

$\vert 1 R\rangle \vert 2L \rangle$ particle 1 is in the right input port and
particle 2 is in the left input port.

Since identical quantum particles are ``indistinguishable", the following input state is
equally possible :
$\vert 1 L\rangle \vert 2 R \rangle$ particle 1 is in the left input port and
particle 2 is in the right input port.

Most general input state is thus constructed as the linear superposition of the two :
$
\Psi= c_1 \vert 1 R\rangle \vert 2L \rangle + c_2 \vert 1 L\rangle \vert 2 R \rangle \;,
$
where $\vert c_1\vert^2+  \vert c_2\vert^2=1$ if two states are orthogonal.
A physical state which represents a real system consisting of identical quantum particles is
either symmetric or anti-symmetric with respect to the permutation of any two particles (the experimental fact);
bosons:
\begin{equation}
\label{SEfu}
\Psi= \frac{1}{\sqrt{2}} (\vert 1 R\rangle \vert 2L \rangle +  \vert 1 L\rangle \vert 2 R \rangle) \;,
\end{equation}
fermions:
\begin{equation}
\label{SEfu}
\Psi= \frac{1}{\sqrt{2}} (\vert 1 R\rangle \vert 2L \rangle - \vert 1 L\rangle \vert 2 R \rangle) \;,
\end{equation}
The QM formalism predicts that the interaction of a bi-boson system with the BS produces
the {\it bunching effect};  in the fermionic case we obtain the {\it anti-bunching effect.}
Our aim is to model these effects in the framework of the classical signal theory. We shall see
that prequantum bi-signals with permutation symmetric components produce bunching and
with anti-symmetric components -- anti-bunching.

\section{Interaction of classical signal with BS}
\label{bunch1}

Consider a classical random signal $\chi$ interacting with BS; one
part of $\chi$ goes to the left input port  of BS, it is denoted
$\chi_L,$ another goes to the right input port  of BS, it is
denoted $\chi_R.$ Hence,$\chi=(\chi_L, \chi_R).$ From classical
electrodynamics we know that, for the input-signal
$\chi_{\rm{in}}= (\chi_{\rm{in} L}, \chi_{\rm{in} R}),$ the output
signal $\chi_{\rm{out}}= (\chi_{\rm{out} L}, \chi_{\rm{out} R}),$
where now indexes $L$ and $R$ are used for output ports of BS, can
be obtained as
\begin{equation}
\label{SE2jip}
\chi_{\rm{out}} (\omega)= U \chi_{\rm{in}}(\omega),
\end{equation}
where, for 50\% -- 50\% BS, $U$ is rotation by the angle $\pi/4$ (the classical Malus law):
 \begin{equation}
\label{COj} U =\frac{1}{\sqrt{2}}\left( \begin{array}{ll}
 1 &  -1 \\
 1 & \; \; 1
 \end{array}
 \right ).
\end{equation}
Now take a bi-signal $$\phi_{\rm{in}}= (\phi_{\rm{in}1},
\phi_{\rm{in}2});$$  each of its component $\phi_{\rm{in}i}=
(\phi_{\rm{in}i L}, \phi_{\rm{in}i R}),i=1,2.$ is transformed by
the Malus law (\ref{COj}). Hence, the output bi-signal is given by
\begin{equation}
\label{SE2jip1} \phi_{\rm{out}}(\omega)= U \times U
\phi_{\rm{in}}(\omega),
\end{equation}
see (\ref{SE2ji2}). We are interested in the off-diagonal block of the covarince operator of
$\phi_{\rm{out}}.$  By (\ref{SE2ji3}) this is $\widehat{\Psi}_{\rm{out}},$ where
$\Psi_{\rm{out}}= U\otimes U \Psi_{\rm{in}}.$
Hence, for a bosonic input bi-signal, i.e., with $\Psi_{\rm{in}}$ given by (\ref{SEfu}), we obtain
\begin{equation}
\label{SEfu2}
\Psi_{\rm{out}} = \frac{1}{\sqrt{2}} (\vert 1 R\rangle \vert 2R \rangle -  \vert 1 L\rangle \vert 2 L \rangle) \;,
\end{equation}
For a fermionic input bi-signal, i.e., with $\Psi_{\rm{in}}$ given by (\ref{SEfu}), we obtain
\begin{equation}
\label{SEfu3} \Psi_{\rm{out}} = \frac{1}{\sqrt{2}} (\vert 1
R\rangle \vert 2L \rangle -  \vert 1 L\rangle \vert 2 R \rangle).
\end{equation}
We now show that the off-diagonal block
$\widehat{\Psi}_{\rm{out}},$ where
 $\Psi_{\rm{out}}$ is given by (\ref{SEfu2}), implies bunching of classical signals;
(\ref{SEfu3}) implies anti-bunching. We consider the later case, i.e., a fermionic bi-signal. Set
$\phi \equiv \phi_{\rm{out}}.$

The intensities of signal's components $(i=1,2)$ at the points $x=R.L$  are given by $I_{i x}=\vert
\phi_{i x} \vert^2$   and their covariations are
\begin{equation}
\label{MU}
g_{xy}=\langle (I_{1x} - \langle I_{1x} \rangle)  (I_{2y} -
\langle I_{2y} \rangle)\rangle,
\end{equation}
where $x, y=R, L.$

We introduce the operators (projectors)  $\widehat{A}_x=\vert
x\rangle \langle x\vert, x=R, L.$ Then $\vert \phi_{i x} \vert^2 =
\langle \widehat{A}_x \phi_{i x}, \phi_{i x} \rangle, i=1,2.$
(These are intensities of bi-signal's components, $i=1,2,$ at the output ports $x=R, L.)$

Hence, by the basic equality (\ref{01K}) we have
\begin{equation}
\label{SEfu5} g_{RR} = \langle \widehat{A}_R \otimes \widehat{A}_R
\Psi_{\rm{out}}, \Psi_{\rm{out}}\rangle,\;  g_{RL} = \langle
\widehat{A}_R \otimes \widehat{A}_L \Psi_{\rm{out}},
\Psi_{\rm{out}}\rangle.
\end{equation}
By taking fermionic $\Psi_{\rm{out}},$ see (\ref{SEfu3}), we
obtain
\begin{equation}
\label{SEfu5q} g_{RR}=0 <g_{RL} =1/2.
\end{equation}
This is the anti-bunching phenomenon. By similar considerations in
the bosonic case we obtain
\begin{equation}
\label{SEfu5q} g_{RL}=0 <g_{RR} =1/2.
\end{equation}
This is the bunching phenomenon.

In fact, this is not surprising. Since we coupled classical field
correlations with quantum correlations, see (\ref{01K}), we can
expect that all quantum effects will be reproduced.

\section{Vector-valued prequantum fields, internal degrees of freedom}

Set $H_{\rm{space}}= L_2({\bf R}^3),$ the space of complex-valued functions which 
are square summable with respect to the Lebesgue measure $dx=dx_1dx_2dx_3;$ set
$H_{\rm{internal}}={\bf C}^n,$ the space of internal degrees of freedom. Consider now
vector fields $\phi(x)=(\phi^{1}(x),..., \phi^{n}(x)),$ where each coordinate field belongs to 
$H_{\rm{space}}.$ Hence, $\phi \in L_2({\bf R}^3;{\bf C}^n),$ the space of vector-valued square summable functions
(with the square of the norm $\Vert \phi \Vert^2=\int  (\sum_j \vert \phi^{´j}(x)  \vert^2) dx).$
We remark that, in fact,
\begin{equation}
\label{Sb}
H_1=L_2({\bf R}^3;{\bf C}^n) = H_{\rm{space}} \otimes H_{\rm{internal}}.
\end{equation}
Thus any normalized vector $\Psi \in H_{\rm{space}} \otimes H_{\rm{internal}}$ determines a 
prequantum random vector-field with the covariance operator $D= \vert \Psi \rangle \langle   \Psi\vert.$ 

Consider now a random bi-signal  with components $$\phi_i(x)=(\phi_i^{1}(x),..., \phi_i^{n}(x)), i=1,2,$$
corresponding to a pure quantum state $\Psi \in   L_2({\bf R}^3;{\bf C}^n)\otimes L_2({\bf R}^3;{\bf C}^n).$ 
Mathematically this state can be considered as a vector belonging to the tensor product
\begin{equation}
\label{Sb0}
H_{12} = H_{\rm{space}} \otimes H_{\rm{internal}} \otimes  H_{\rm{space}} \otimes H_{\rm{internal}}.
\end{equation}
Moreover, mathematically it may be convenient to separate the space and internal coordinates and to represent 
$H_{12}$ in the form:  
\begin{equation}
\label{Sb1}
H_{12} = (H_{\rm{space}} \otimes  H_{\rm{space}}) \otimes   (H_{\rm{internal}}  \otimes H_{\rm{internal}}).
\end{equation}
We remark that this is just a mathematical representation´.

This construction can be generalized to the space $L_2(X, \nu),$ where $X$ is a set and $\nu$ is a measure 
on this set; here $\Vert \phi \Vert^2= \int_X \vert \phi(x) \vert^2 d\nu(x).$  In fact, we are interested 
in a simple case: $X=\{x_1,...,x_m\}$ is a finite set and $\nu$  is the uniform measure on $X,$ i.e.,
$\nu(x)=1$ for any $x.$ Hence, instead of the integral, we have the sum
$\Vert \phi\Vert^2= \sum_k \vert \phi(x_k)\vert^2.$ So, the space is discrete and $H_{\rm{space}}= {\bf C}^m.$ 
By taking into account the internal degrees of freedom we obtain tensors $\phi= (\phi^{ji}),  j=1,...,m, i=1,...,n.$
The first index describes the spatial degrees of freedom and the second describes internal ones.
Thus (\ref{Sb})   has the form:
\begin{equation}
\label{Sb2}
H_1= {\bf C}^m \otimes {\bf C}^n.
\end{equation}
For a bi-signal, we have
\begin{equation}
\label{Sb3}
H_{12}= ({\bf C}^m \otimes {\bf C}^n)\otimes ({\bf C}^m \otimes {\bf C}^n)) .
\end{equation}
In purely, mathematical consideration we can work with the space
\begin{equation}
\label{Sb4}
H_{12}= ({\bf C}^m \otimes {\bf C}^m)\otimes ({\bf C}^n \otimes {\bf C}^n) .
\end{equation}

\section{Bunching and anti-bunching: prequantum fields with spin}
\label{bunch2}

In our formalism  spin $1/2$ quantum system is described in the following way.

Take  $X=\{R,L\},$ i.e., $m=2;$ set $n=2$ (two internal degrees of freedom):
$H_{\rm{space}}={\bf C}^2, H_{\rm{internal}}={\bf C}^2.$
 We obtain 
${\bf C}^2$-valued ``fields''  $\phi=(\phi(R)^{+}, \phi(R)^{-}, \phi(L)^{+}, \phi(L)^{-}).$ 
Such random vector describes  spin $1/2$ quantum system. Take, for example, 
$\psi=\vert R\rangle \vert +\rangle.$ It determines a random   spin $1/2$  
prequantum field.  Now consider two such fields, a bi-signal with components
$\phi=(\phi_i^{+}(R), \phi_i^{-}(R), \phi_i^{+}(L), \phi_i^{-}(L)), i=1,2.$
Consider the vector
\begin{equation}
\label{Sb5}
 \Psi= \frac{1}{2} (\vert 1 R\rangle \vert 2L \rangle - \vert 1 L\rangle \vert 2 R \rangle )\otimes 
(\vert +\rangle_1 \vert - \rangle_2 - \vert -\rangle_1 \vert+ \rangle_2).   
\end{equation}
This is the purely mathematical representation, see (\ref{Sb4}); to proceed to PCSFT, we rewrite 
 $$
\Psi= \frac{1}{2} (\vert 1 R\rangle \vert +\rangle_1 \otimes  \vert 2L \rangle\vert -\rangle_2 - 
\vert 1 R\rangle \vert -\rangle_1 \otimes  \vert 2L \rangle\vert +\rangle_2 
$$
\begin{equation}
\label{Sb51}
- \vert 1 L\rangle \vert +\rangle_1 \otimes  \vert 2R \rangle\vert -\rangle_2 +
\vert 1 L\rangle \vert -\rangle_1 \otimes  \vert 2R \rangle\vert +\rangle_2).
\end{equation}
This vector $\Psi \in (H_{\rm{space}} \otimes {\bf C}^2) \otimes (H_{\rm{space}} \otimes {\bf C}^2)$
determines the off-diagonal block of the covariance matrix of the Gaussian bi-signal with componens
valued in the space $H_{\rm{space}} \otimes {\bf C}^2,$ the space of fields from $H_{\rm{space}}$ to ${\bf C}^2.$

The intensities of signal's components  $(i=1,2)$ at the points $x=R.L$ are given by 
$$I_{i x}=\vert \phi_i^{+}(x) \vert^2 + \vert \phi_i^{-}(x) \vert^2$$   and their covariations are
$$
g_{xy}=\langle (I_{1x} - \langle I_{1x} \rangle)  (I_{2y} -
\langle I_{2y} \rangle)\rangle,
$$
where $x, y=R, L.$

We introduce the operators (projectors)  $\widehat{A}_{1x}=\vert
x\rangle \langle x\vert \otimes I$ and 
$\widehat{A}_{2x}=  I\otimes \vert
x\rangle \langle x\vert,$ where  $x=R, L.$
The corresponding quadratic forms (physical variables of PCSFT) are given by
$$
f_{A_{ix}}(\phi_i) = \vert \phi_i^{+}(x) \vert^2 + \vert \phi_i^{-}(x) \vert^2 =
I_{i x}, \; i=1,2, x=R.L.
$$
As in the case of spinless fields, see (\ref{MU}), we define covariations of intensities.
By the basic equality coupling prequantum and quantum correlations we obtain
$$
g_{RR}= \langle (f_{A_{1R}}  - \langle f_{A_{1R}}\rangle)  
(f_{A_{2R}}  - \langle f_{A_{2R}} \rangle) \rangle =
\langle \widehat{A}_{1R} \otimes \widehat{A}_{2R} \Psi, \Psi \rangle 
$$
\begin{equation}
\label{BCC}
=\langle \widehat{A}_{R} \otimes \widehat{A}_{R} \Psi_{\rm{space}}, \Psi_{\rm{space}} \rangle \Vert \Psi_{\rm{spin}} \Vert^2 =0,
 \end{equation}
 where $\Psi_{\rm{space}}= \frac{1}{\sqrt{2}} (\vert 1 R\rangle \vert 2L \rangle - \vert 1 L\rangle \vert 2 R \rangle )$
 and $\frac{1}{\sqrt{2}}(\vert +\rangle_1 \vert - \rangle_2 - \vert -\rangle_1 \vert+ \rangle_2).$   
 (We state again that the latter vectors are purely mathematical expressions; physically space and spin components 
 are coupled.)
 
Hence, in the complete accordance with formal predictions of QM the classical bosonic (permutation symmetric) spin  1/2 bi-signals
 feature ``fermionic collision'': $g_{RR}=0.$ 
If we consider a fermionic   spin  1/2 bi-signal such that the inter-components correlations are given by the operator 
$\widehat{\Psi}$ corresponding to the vector which can be formally written as
\begin{equation}
\label{Sb5k}
 \Psi= \frac{1}{2} (\vert 1 R\rangle \vert 2L \rangle + \vert 1 L\rangle \vert 2 R \rangle )\otimes 
(\vert +\rangle_1 \vert - \rangle_2 - \vert -\rangle_1 \vert+ \rangle_2),
\end{equation}
then again in the complete accordance with formal predictions of QM such bi-signal
 features ``bosonic collision'': $g_{RL}=0.$

{\bf Conclusion.} {\it Bunching and anti-bunching for quantum
systems can be represented in the classical signal framework.}

\end{document}